\documentclass[twocolumn,showpacs,aps,prl]{revtex4}

\newcommand{\bk}{{\bf k}}

\newcommand{\beq}{\begin{eqnarray}}
\newcommand{\eeq}{\end{eqnarray}}
\newcommand{\beqq}{\begin{eqnarray*}}
\newcommand{\eeqq}{\end{eqnarray*}}

\usepackage{graphicx}
\usepackage{dcolumn}
\usepackage{bm}
\usepackage{color}

\begin{document}

\begin{titlepage}

\title{Spin-filtered Edge States with an Electrically Tunable Gap in a Two-Dimensional Topological Crystalline Insulator}

\author{Junwei Liu$^{1,2}$, Timothy H. Hsieh$^2$, Peng Wei$^{2,3}$, Wenhui Duan$^1$, Jagadeesh Moodera$^{2,3}$ and Liang Fu$^{2}$}
\email{liangfu@mit.edu}
\address{$^1$Department of Physics and State Key Laboratory of Low-Dimensional Quantum Physics, Tsinghua University, Beijing 100084, People's Republic of China\\ $^2$Department of Physics, Massachusetts Institute of Technology, Cambridge, MA 02139 \\
$^3$Francis Bitter Magnet Lab, Massachusetts Institute of Technology, Cambridge, MA 02139}

\begin{abstract}
Three-dimensional  topological crystalline insulators were recently predicted and observed in the SnTe class of IV-VI semiconductors, which host metallic surface states protected by crystal symmetries.  In this work, we study thin films of these materials and expose their potential for device applications.  We demonstrate that  thin films of SnTe and Pb$_{1-x}$Sn$_{x}$Se(Te) grown along the (001) direction are topologically nontrivial in a wide range of film thickness and carry conducting spin-filtered edge states that are protected by the (001) mirror symmetry via a topological invariant.  Application of an electric field perpendicular to the film will break the mirror symmetry  and generate a band gap in these edge states.  This functionality motivates us to propose a novel topological transistor device, in which charge and spin transport are maximally entangled and simultaneously controlled by an electric field.  The high on/off operation speed and coupling of spin and charge in such a device may lead to electronic and spintronic applications for topological crystalline insulators.
\end{abstract}

\maketitle

\draft

\vspace{2mm}

\end{titlepage}

{\bf Crystal structure and symmetry play a fundamental role in determining electronic properties of quantum materials.
The interplay between
crystallography and electronic topology\cite{fukane, teofukane} has advanced our understanding of topological insulators\cite{kanehasan, zhangreview, moore}. More recently, a new type of topological phases termed topological crystalline insulators\cite{fu} has been predicted\cite{hsieh} and  observed\cite{ando, poland, hasan} in three-dimensional materials SnTe and Pb$_{1-x}$Sn$_x$Se(Te). A key characteristic of topological crystalline insulators is the presence of metallic boundary states that are protected by crystal symmetry, rather than time-reversal\cite{mong}. As a consequence, these states can acquire a band gap under perturbations that break the crystal symmetry\cite{hsieh, vidya}. This novel property opens up an unprecedented functionality of tuning the charge and spin transport of topological boundary states  with high on/off speed by applying an electric field. Here we theoretically demonstrate that thin films of SnTe and Pb$_{1-x}$Sn$_x$Se(Te) grown along the (001) direction realize a new, two-dimensional topological crystalline insulator phase that supports spin-filtered edge states with  a band gap tunable by electric field effect. Our work may thus enable electronic and spintronic device applications based on topological crystalline insulators. }

Topological crystalline insulators (TCI) have so far only been realized in three-dimensional materials\cite{hsieh, ando, poland, hasan}. In this work, we propose a material realization of a two-dimensional (2D) topological crystalline insulator phase, which is topologically distinct from an ordinary insulator in the presence of mirror symmetry about the 2D plane. The topology here is mathematically characterized by two integer topological invariants $N_+$ and $N_-$, which are Chern numbers of Bloch states with opposite mirror eigenvalues. While the sum $N_+ + N_-$ determines the quantized Hall conductance, the mirror Chern number\cite{teofukane} defined by $N_M \equiv (N_+ - N_- )/2$ serves as an independent topological index\cite{hsieh, murakami, fiete, yao, furusaki, sun, dai}, which distinguishes a mirror-symmetric TCI in two dimensions.

The mirror eigenvalue of an electron wavefunction is intimately related to its spin.
Because performing mirror transformation (denoted by $M$) twice is equivalent to a $2\pi$ rotation which changes the sign of a spinor,
$M^2=-1$ and hence mirror eigenvalues are either $i$ or $-i$.
In the absence of spin-orbit coupling, these two eigenvalues correspond to spin eigenstates $s_z=\pm \frac{1}{2}$ where the quantization axis $z$ is perpendicular to the plane, so that mirror Chern number $N_M$ reduces to spin Chern number\cite{lee, young, kim}.
More broadly, mirror Chern number can be defined for any system that is symmetric under reflection, with or without spin conservation.
For instance, the quantum spin Hall state in graphene proposed by Kane and Mele\cite{km} is characterized by $|N_M|=1$, in addition to the
$Z_2$ index.

In this work, we demonstrate that (001) thin films of topological crystalline insulator SnTe and Pb$_{1-x}$Sn$_{x}$Se(Te) in a wide range of thickness realize a  2D topological phase indexed by mirror Chern number $|N_M|=2$, which supports {\it two} pairs of spin-filtered edge states. This new  topological phase belongs to a different symmetry class than the quantum spin Hall state, and its edge states are protected solely by mirror symmetry instead of time reversal.
A {\it unique} consequence is that applying an electric field perpendicular to the film breaks the mirror symmetry and generates a band gap in these edge states. This functionality leads to a new way of controlling the charge and spin transport simultaneously by electric field effect in a two-dimensional topological crystalline insulator.

Our results are based on a combination of $k\cdot p$ modeling, band structure calculations, and topological band theory.  A detailed derivation of the following analysis can be found in Methods.  Here, we describe how the competition between (1) the inverted band structure in the three-dimensional limit and (2) hybridization between the two surfaces determines
the topological nature of the TCI film as a function of film thickness. SnTe and Pb$_{1-x}$Sn$_x$Te films are typically grown by molecular beam epitaxy in a layer-by-layer mode with good thickness control\cite{mbe1, mbe2, mbe3}.

Three-dimensional TCI, SnTe or Pb$_{1-x}$Sn$_{x}$Se(Te), have Dirac fermion surface states on the (001) surface.  There are four Dirac points located at non-time-reversal-invariant momenta $\Lambda$ near the $X$ points, which can be derived from the $k\cdot p$ theory at the $X$ point \cite{kp}.  Surface states at $X$ correspond to $p$-orbitals of the cation (C) and the anion (A) respectively. Importantly, TCI has an inherently inverted band ordering opposite to the ionic insulator PbTe\cite{hsieh}, so that the anion-derived surface state $E_A(X)$ is located near the conduction band edge, and the cation-derived surface state $E_C(X)$ near the valence band edge\cite{kacman}.  
As we will show below,  the topological properties of TCI (001) films are dictated by these gapped surface states at $X$, while their low-energy band structures are largely determined by the gapless states at $\Lambda$.

\begin{figure}[tbp]
\includegraphics[width=8.5cm]{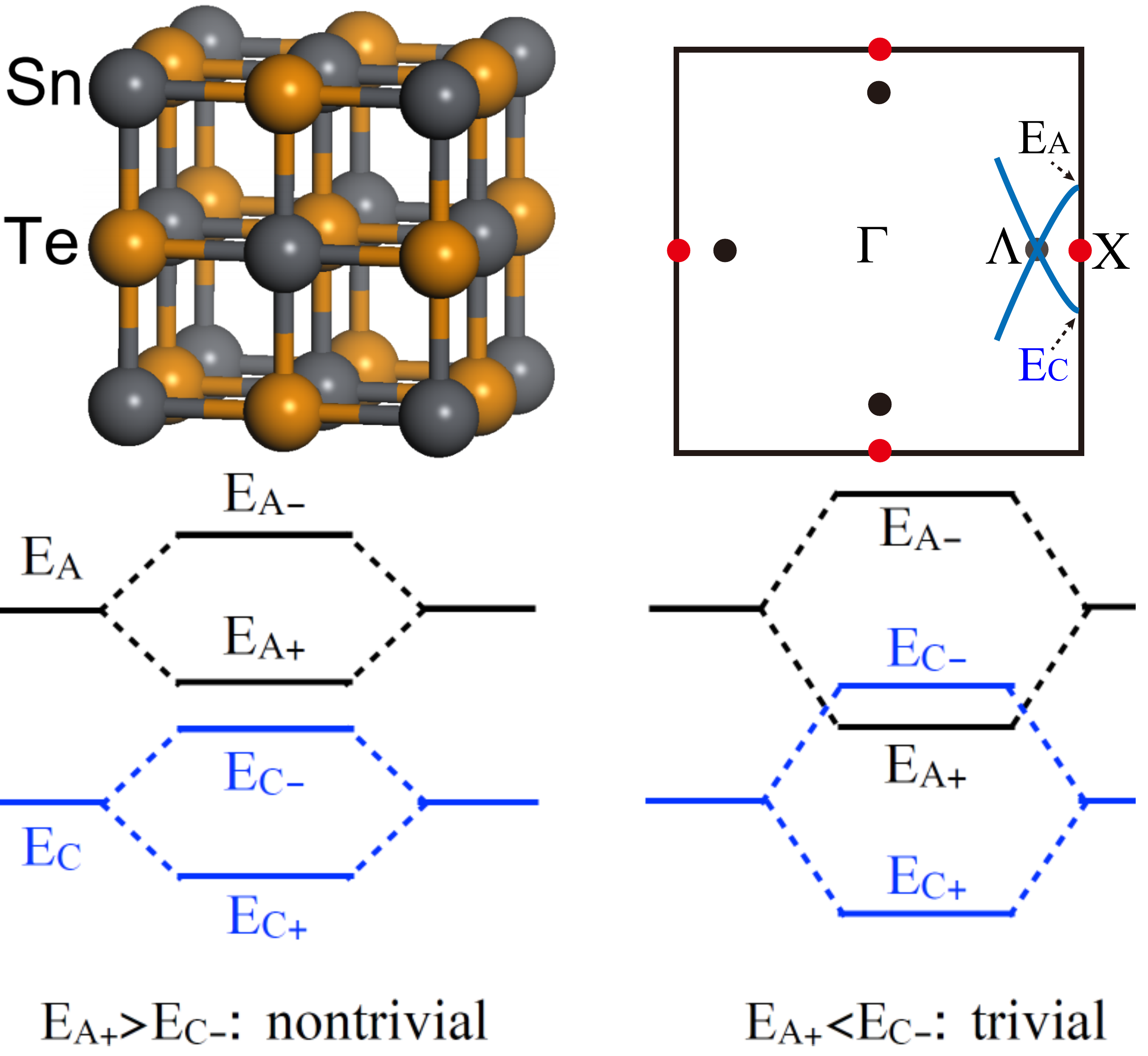}
\caption{{\bf  Energy Level Diagram of TCI Thin Films.} 
Top: rocksalt structure of SnTe (left) and schematic Brillouin zone of the (001) surface of three-dimensional SnTe (right). Four massless Dirac points of surface states are indicated by the black dots. Bottom: schematic conduction and valence bands of a TCI film at the $X$ point.  When the top and bottom surfaces hybridize weakly, the film inherits the inverted band gap of the 3D limit (left).  Strong hybridization drives a band crossing, leading to a trivial phase (right). $A, C$ denote anion/canion, and $\pm$ denote bonding/anti-bonding combinations of the two surfaces.}
\end{figure}

In this work, we consider TCI (001) films with an odd number of atomic layers, which are symmetric under the reflection $z\rightarrow -z$ about the 2D plane in the middle (see Fig.1). 
Our main result---gapless edge states with an electrically tunable gap---also holds for TCI films with an even number of layers, although the underlying symmetry is different and to be described elsewhere. 
When the film thickness is below the penetration length of surface states,
the wavefunction hybridization between the top and bottom surfaces results in an energy splitting between the bonding and anti-bonding states.
These hybridized surface states form the conduction and valence bands of the 2D TCI film.
Based on $k\cdot p$ analysis (see Methods) and band structure calculations, we find that the conduction and valence bands of the TCI film at the $X$ point
are derived from the bonding state of the anion at energy $E_{A+}(X)$ and the anti-bonding state of the cation at energy $E_{C-}(X)$.
Due to their opposite parity, the two bands $E_{A+}(X)$ and $E_{C-}(X)$ do not repel each other.
The band ordering of   $E_{A+}(X)$ and $E_{C-}(X)$ depends crucially on the competition between the
hybridization of the two surfaces and the inverted gap $2m$ of each surface.
For thick films, the hybridization is weak so that $E_{A+}(X)>E_{C-}(X)$. With this ordering, the band structure of the TCI film is
adiabatically connected to the original surface states in the three-dimensional limit, which is inherently inverted near $X$ points (see Fig.1).

We now concentrate on the band inversion near the transition point $E_{A+}(X)\sim E_{C-}(X)$, which occurs at a critical film thickness to be determined from band structure calculations later. This transition is described by Dirac mass reversal in a $k\cdot p$ Hamiltonian:
\beq
H(\bk) = (  \tilde{v}_x k_x  s_x   - \tilde{v}_y k_y s_y ) \tau_x +  \tilde{m}  \tau_z
\eeq
where $\tau_z=\pm 1$ denotes the conduction and valence band of the TCI film at $X$, and each band has a Kramers degeneracy labelled by  $s_z$.
The velocities $\tilde{v}_x, \tilde{v}_y$ and the Dirac mass $\tilde{m} \equiv E_{C-}(X) - E_{A+}(X) $ are derived from
microscopic parameters of surface states in the 3D TCI and their hybridization strengths (see Methods for details).

Importantly, $H(\bk)$ is invariant under mirror symmetry about the 2D plane:
\beq
M H(\bk) M^{-1} = H(\bk).
\eeq
Here the mirror operator is given by $M= -i s_z \tau_z$, because bonding and anti-bonding states have opposite mirror eigenvalues, and so do spin up and down states.
For a single-flavor Dirac fermion,
sign reversal of Dirac mass $\tilde{m}$ changes mirror Chern number $N_M$ by $1$\cite{teofukane}.
We further take into account that  there exist two $X$ points in the Brillouin zone,
related by a four-fold rotation. Because of the simultaneous band inversions at both $X$ points,
 $N_M$ changes by $2$:
\beq
|N_M(\tilde{m}>0)  - N_M(\tilde{m}<0)|= 2.
\eeq
Combining this equation with the inverted band structure of $\tilde{m}<0$ deduced earlier,
we conclude that (001) thin films of  TCI with an inverted band ordering $E_{A+}(X)>E_{C-}(X)$
realize a two-dimensional topological crystalline insulator phase with mirror Chern number $|N_M|=2$.

\begin{figure}[tbp]
\includegraphics[width=8cm]{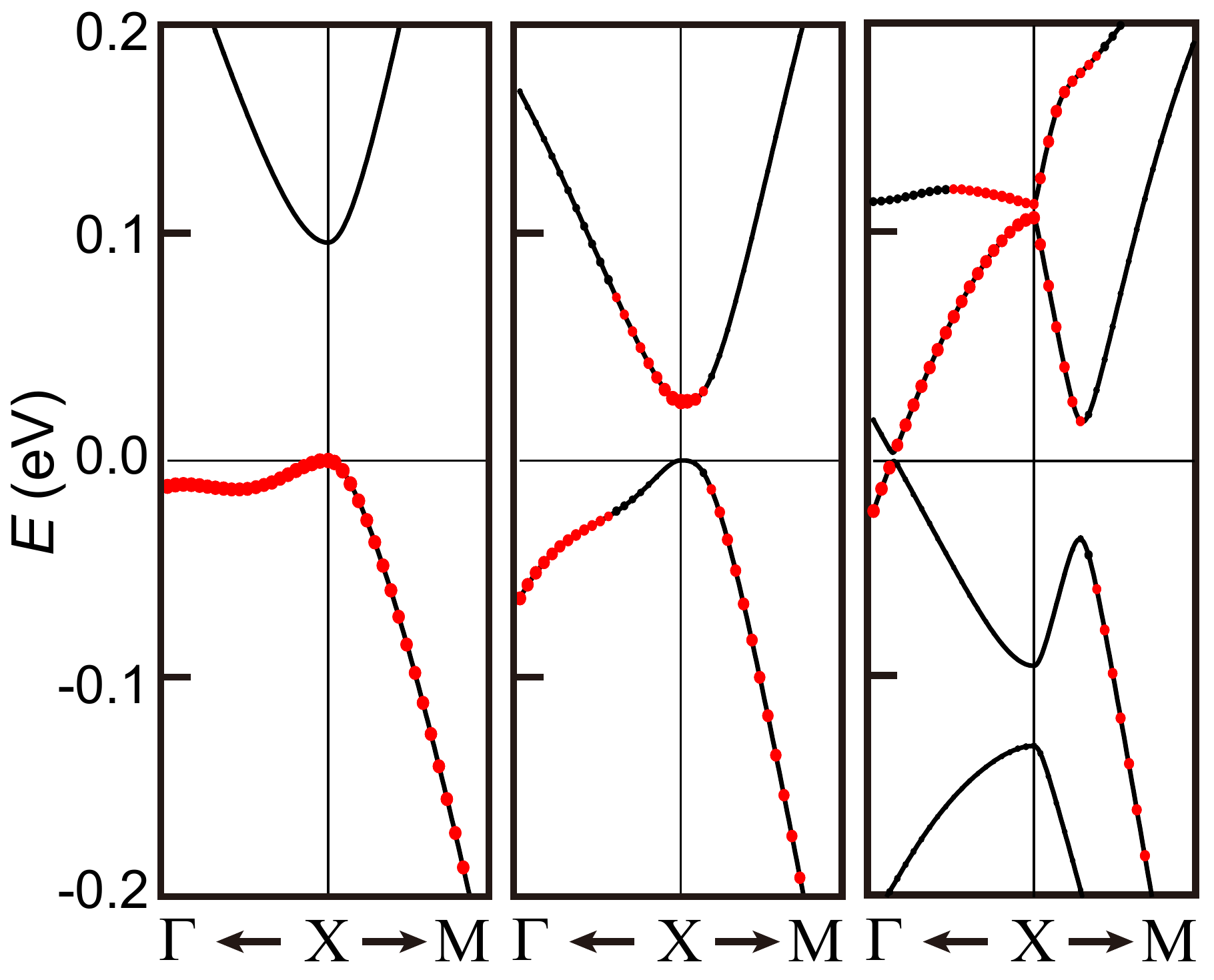}
\caption{{\bf Band Inversion in SnTe Films.} Conduction and valence bands of a 3 layer (left), 5 layer (middle), and 25 layer (right) SnTe  film.  Red dots denote the weight of the electronic wavefunction on the anion Te. 
Starting from 5 layers, TCI (001) films show an inverted band structure relative to an ionic insulator. }
\end{figure}

The above conclusion drawn from $k\cdot p$ analysis and topological band theory
is confirmed by our band structure calculations of (001) TCI films based on a realistic tight-binding model\cite{lent}, using SnTe as a representative. Fig.2 shows the band structures and orbital characters near the $X$ point for 3- and 5-layer films, as well as a thick 25-layer film.
Clearly, the 5-layer film has an inverted band ordering with the conduction (valence) band near $X$ derived from the anion (cation), while the 3-layer film does not. This confirms our $k\cdot p$ result that the increased
hybridization strength in thinner films drives the system out of the inverted regime. We further calculate the mirror Chern number numerically by integrating the Berry curvature over the Brillouin zone (see Methods), and confirm that $N_M=0$ for the 3-layer film and $|N_M|=2$ for the 5-layer film.

As the film thickness increases above 5 layers, the band gap at $X$, $E_g(X) \equiv E_{A+}(X)-E_{C-}(X)$ increases monotonically, and eventually reaches the value $220$meV 
that is determined by the band structure of surface states in the 3D limit. Meanwhile, the fundamental band gap in thicker films shifts from $X$ to
the momentum $\Lambda$ located on the line $\Gamma X$, see for example the band structure of the 25-layer film shown in Fig.2.  
In the 3D limit, the gap at $\Lambda$ decreases to zero, thus recovering zero-energy  Dirac points of TCI surface states (see Fig.3). 
Irrespective of this shift of band gap, films thicker than 5 layers all show an inverted band structure near the $X$ point, which results in the nontrivial band topology. 
This leads to a robust 2D topological crystalline insulator phase in a {\it wide} range of  thickness. 
Moreover, the inverted gap in 11-layer SnTe film reaches $0.15$eV near the $X$ point,  which is much larger than that of existing quantum spin Hall insulators.
The same physics holds for Pb$_{1-x}$Sn$_x$Se and Pb$_{1-x}$Sn$_x$Te, though the critical thickness depends on material details.

\begin{figure}[tbp]
\includegraphics[width=8cm]{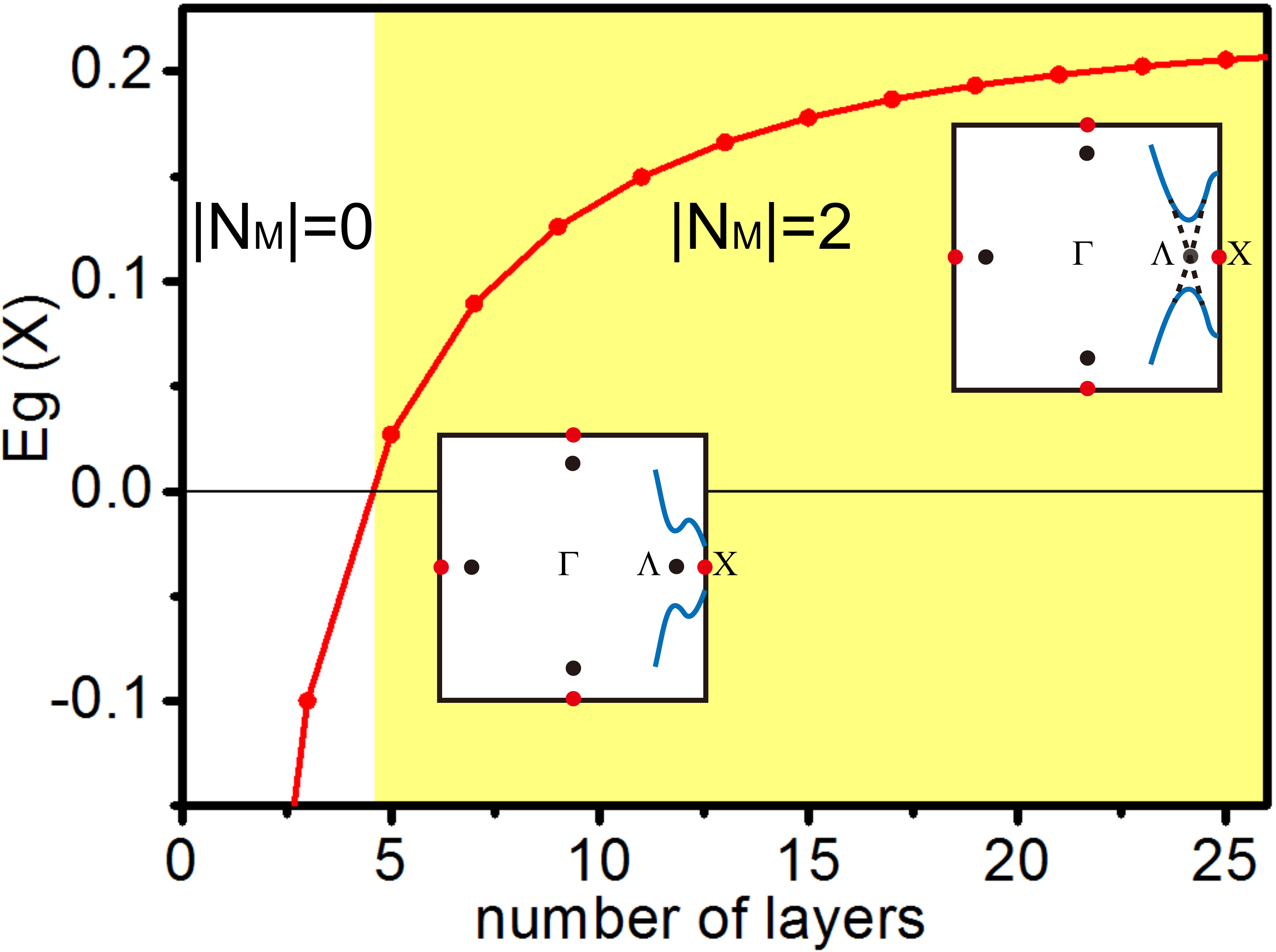}
\caption{{\bf Phase Diagram of 2D TCI.} Band gap of SnTe at the $X$ point as a function of film thickness.  Above five layers, the gap of SnTe at $X$ is inverted and increases with thickness, resulting in a wide region of topologically nontrivial 2D TCI phase with mirror Chern number $|N_M|=2$. For thick films, the fundamental band gap shifts from $X$ to $\Lambda$, where the Dirac points of 3D surface states are located (insets). }
\end{figure}

A hallmark of 2D TCI is the presence of conducting edge states.  The mirror Chern number $|N_M|=2$ dictates that there exist two pairs of counter-propagating edge states within the band gap, and those moving in the same (opposite) direction carry identical (opposite) mirror eigenvalues. This is confirmed in our band structure calculation of a SnTe thin film in a strip geometry parallel to $[100]$, using the recursive Green's function method\cite{gf}.
As shown in Fig. 4, edge states with opposite mirror eigenvalues cross each other at a pair of momenta  in the edge Brillouin zone.
In the ballistic limit, the conductance through such edge states is $2e^2/h$.

Unlike helical edge states in a quantum spin Hall insulator,  the band crossings of spin-filtered edge states found here are located at {\it non}-time-reversal-invariant momenta, so that they are protected solely by the mirror symmetry $z \rightarrow -z$, but not time-reversal.
This leads to a remarkable consequence: applying a perpendicular electric field, which breaks the mirror symmetry, will
 generate a band gap in these edge states.
To  illustrate this effect and estimate its magnitude, we calculate the band structure of an 11-layer SnTe film under
a modest, uniform electric field  that generates a $0.1$eV potential difference across the film. 
The effect of an external electric field is modeled by adding a layer-dependent potential that 
varies linearly in the (001) direction to the tight-binding model\cite{lent}.   
We find that
the spin-filtered edge states become completely gapped (see Fig.4).
For comparison, a magnetic field or  induced ferromagnetism is required to gap helical edge states of  a quantum spin Hall insulator, which can be 
difficult to achieve\cite{du}.

\begin{figure}[tbp]
\includegraphics[width=8cm]{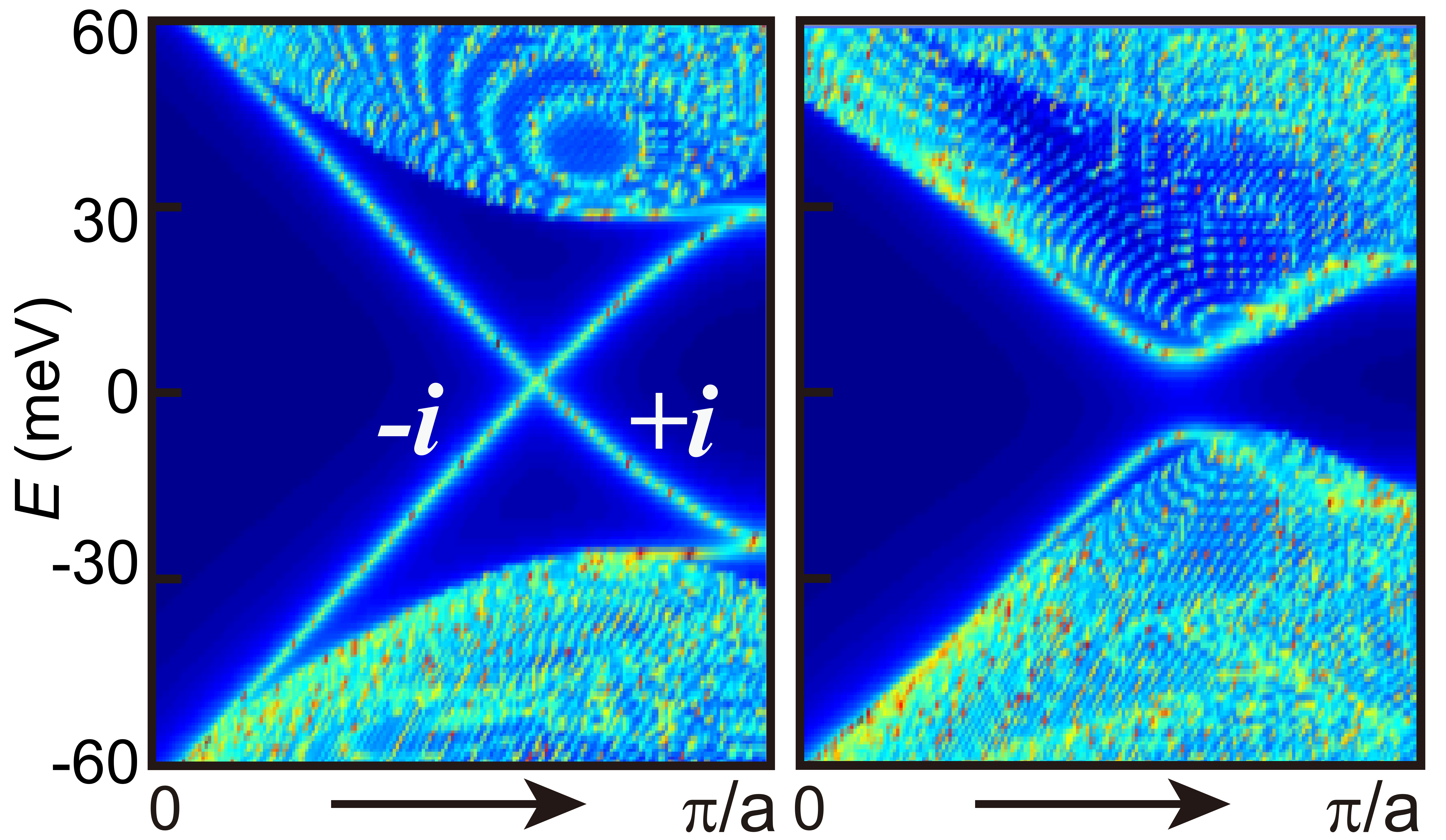}
\caption{{\bf Effect of Electric Field on Edge States.} Edge states of a TCI film and electric-field-induced gap.  (Left) Bulk and gapless edge dispersion of a 11 layer SnTe film, labeled by mirror eigenvalues. (Right) Gapped dispersion when a perpendicular electric field is applied to generate a potential difference of 0.1eV across the film.}
\end{figure}

Thus edge-state transport in the 2D topological crystalline insulator phase found here has the unique property that the conductance is easily and widely tunable via an electric-field-induced gap instead of carrier depletion. This mechanism works at high on-off speed and improves power efficiency. As the on/off states originate from the crystal symmetry and symmetry breaking, they are more robust to material imperfections, which may also increase the operating cycles and improve the performance of the transistor especially at high frequency. 
In addition, only a local electric field is required to gap out edge states, which further minimizes power consumption. 
The resistance of the ``Off'' state depends on the amount of impurity states inside the energy gap, and thus the On/Off ratio can be improved by controlling the film quality\cite{onoff}. On the other hand, in the ballistic transport regime, the On/Off ratio of the topological transistor can be further enhanced with a quantized "On" state conductance of $2 e^2/h$ per edge and a negligible "Off" state conduction.

Since the electron spin polarization in the $z$ direction is proportional to mirror eigenvalue, an electric current  carried by the TCI edge states is spin-polarized, and reversing the current flips the spin (Fig.5). Importantly, it is the mirror eigenvalue, rather than the amount of spin polarization, that determines spin transport.  For instance, when a TCI is placed adjacent to a ferromagnetic lead, $s_z=+1/2$ electrons in the lead will tunnel into the right moving TCI edge states with mirror eigenvalue $+i$ with $100\%$ certainty, 
if the setup preserves the $z\rightarrow -z$ mirror symmetry. 

The above functionality motivates us to propose a topological transistor device made of dual-gated TCI thin films, as shown in Fig.5. The device "On" state in Fig. 5 itself can be used as a spin diode that enables the electric field control of the electron spin polarization. Using the two gates, one can control the electric field across the film and the carrier density independently, and thus turn on and off the coupled charge and spin transport by purely electrical means functioning as a transistor.  

\begin{figure}[tbp]
\includegraphics[width=8.5cm]{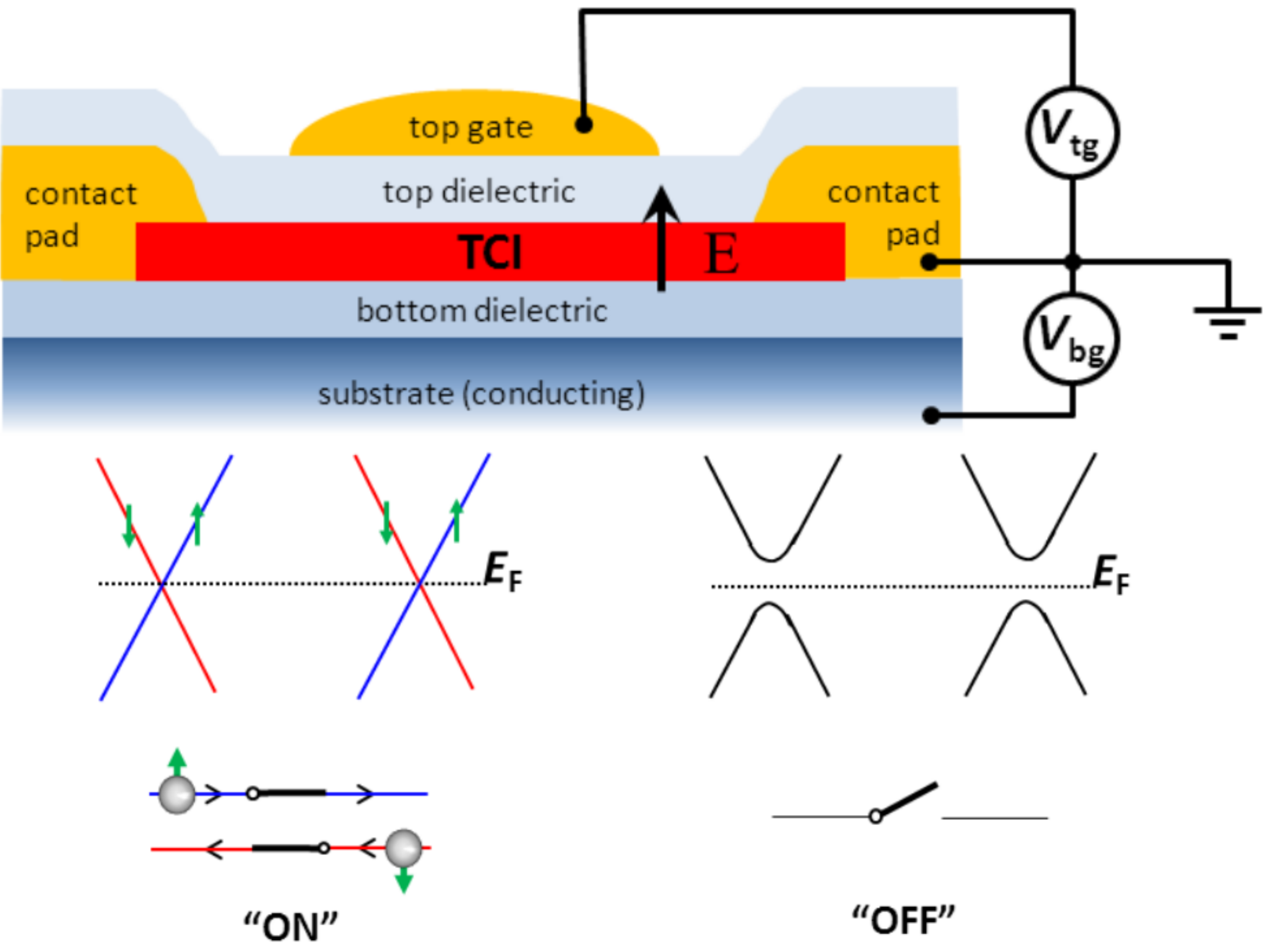}
\caption{{\bf Topological Transistor.} Proposed topological transistor device for using an electric field to tune charge and spin transport.  Without an electric field, the TCI film has mirror symmetry and thus protected, spin-filtered edge states (left).  Applying an electric field perpendicular to the film breaks mirror symmetry, which gaps the edge states (right).}
\end{figure}

In addition to an electric field, a magnetic field can also be applied to a TCI thin film to independently control the dispersion of the two sets of edge modes.  Consider the (100) edge states with crossings at $\pm k_0$, and apply a magnetic field $B=(B_{||}, B_{\perp}, B_z)$ (parallel being along the edge).  $B_z$ will shift both the energies and momenta of these crossings, 
while in-plane magnetic field $B_{||}$ and $B_z$ open up gaps. 
By tuning both electric and magnetic fields,  the gaps at $\pm k_0$ can be separately tuned in full range:
\beq
\Delta_{k_0} &=& \sqrt{(m_E + m_{B_{\perp}})^2 +m_{B_{||}}^2} \label{gap} \\
\Delta_{-k_0} &=& \sqrt{(-m_E + m_{B_{\perp}})^2 +m_{B_{||}}^2}.
\eeq  
Here $m_{E}, m_{B_{\perp}}, m_{B_{||}}$ are energy gaps that are generated separately from the corresponding fields (see Methods).

From the material standpoint, thin films of IV-VI semiconductors have been grown epitaxially\cite{mbe1,mbe2,mbe3} and extensively studied\cite{review}. In particular, ultrathin SnTe quantum wells have been made with high quality\cite{qw1,qw2,qw3}. Related materials PbTe, PbSe and PbS, while topologically trivial in bulk form, can become inverted in strained thin films\cite{hsieh}, which further broadens the material base available for use. Quantum devices based on IV-VI semiconductor quantum wells have been fabricated, which show ballistic transport with remarkable conductance quantization\cite{ballistic}. 
These developments as well as intensive interest in TCI thin films \cite{bernevig, kanerecent} give us hope that the two-dimensional topological crystalline insulator phase and topological transistor device proposed here can be experimentally realized in the future.

 {\bf Methods}:
In this section, we present (i) a detailed derivation of $k\cdot p$ theory for TCI film; (ii) computation of mirror Chern number; and (iii) effective theory for TCI edge states under electric and magnetic fields. 

\subsection{I. Derivation of $k \cdot p$ theory for TCI film} 
Here we provide the derivation leading up to the effective Hamiltonian (1).
The starting point is (001) topological surface states of TCI in the three-dimensional limit.
The $k\cdot p$ Hamiltonian for these surface states, linearized at an $X$ point, is given by:
\beq
H_0(\bk)=  ( v_x k_x   s_y + v'_x k_x s_z \tau_x  - v_y k_y s_x ) \sigma_z + m  \tau_z.  \label{001}
\eeq
$\sigma_z=\pm 1$ denote the top/bottom (001) surfaces, $\tau_z=1(-1)$ denote basis states that are mainly derived from the cation(anion), and $s_z=\pm 1$ denotes a Kramers doublet.
For a given value of $\sigma_z$, $H_0(\bk)$ reduces to the four-band $k\cdot p$ Hamiltonian for a single surface,  derived in Ref.\cite{kp, fang, hsin}.  Note, however, for technical convenience, we have chosen a basis different from that in Ref.\cite{kp}, such that $H_0(\bk=0)$ is diagonal.

$H_0(\bk)$ is invariant under the symmetry operations of inversion $P$, $M_z: z \rightarrow -z$, $M_x: x\rightarrow -x$, $M_y: y\rightarrow -y$, two-fold rotation about $z$-axis $C_2$, and time-reversal $\Theta$.  In the above basis, these are represented by
\beq
P&=&\sigma_x, \nonumber\\
M_z &=& -is_z \tau_z \sigma_x, \nonumber \\
M_x&=&-is_x, \nonumber \\
M_y &=& -is_y  \tau_z, \nonumber \\
C_2 &=& -is_z \tau_z, \nonumber \\
\Theta &=&  is_y K.
\eeq

Hybridization between the two surfaces corresponds to off-diagonal terms in $\sigma_z$ basis.  There are three hybridization terms allowed by the above symmetries:
\beq
H_h(\bk) = \delta_1  \sigma_x + \delta_1' \tau_z \sigma_x + \delta_2 s_x \tau_y \sigma_x   . \label{h}
\eeq
Here $\delta_1 \pm \delta_1'$ is the intra-orbital hybridization matrix element within the cation (anion) orbitals on the top and bottom surface.
$\delta_2$ is the inter-orbital hybridization matrix element between the cation and anion orbitals on the two surfaces.
By combining (\ref{001}) and (\ref{h}), we obtain an eight-band $k\cdot p$ Hamiltonian for the (001) thin film of TCI:
\beq
H_t(\bk) = H_0(\bk) + H_h(\bk).
\eeq

The intra-orbital hybridization $\delta_1$ and $\delta_1'$ splits
a pair of degenerate states at $X$ on the two surfaces, $E_{\alpha=C, A}(X)$,  into a bonding and anti-bonding states with
energies $E_{\alpha+}(X)$ and $E_{\alpha-}(X)$ respectively. From
band structure calculations, we find that $\delta_1'$ is smaller than $\delta_1$, so that the bonding states have a lower energy than the anti-bonding states:  $E_{\alpha+}(X) < E_{\alpha-}(X)$. In this case, the conduction and valence band edges at $X$
are derived from the bonding combination of the anions at energy $E_{A+}(X)$ and the anti-bonding orbital of the cations at energy $E_{C-}(X)$.

The transition as these two levels cross each other  can be
derived by projecting $H_t(\bk)$ into the low-energy subspace of $E_{A+}(X)$ and $E_{C-}(X)$.  We find that $\delta'_1$ plays an insignificant role and can be set to zero for simplicity.  Projecting onto the four bands yields the $k\cdot p$ Hamiltonian for TCI film, Eq.(1) 
\beq
H(k) = \tilde{m} \tau_z + ({\tilde v_x} k_x s_x - {\tilde v_y} k_y s_y) \tau_x.
\eeq
where the parameters are given by
\beq
\tilde{m} &=& \sqrt{m^2+\delta_2^2}-|\delta_1|, \nonumber \\
{\tilde v_x}&=&v'_x {\rm sgn}(\delta_1 \delta_2), \nonumber \\
 {\tilde v_y} &=& \frac{v_y |\delta_2|}{\sqrt{m^2+\delta^2_2}}.
 \eeq
 
 \subsection{II. Mirror Chern number}
 
The mirror Chern number $n_M$ is determined by the difference in Chern numbers $N_+$ and $N_-$ that are associated with different mirror eigenstates. We compute 
these Chern numbers by numerically integrating the Berry curvature for all occupied bands over the entire 2D Brillouin zone\cite{fangz}. The Bloch wavefunctions are obtained from 
the tight-binding model for SnTe\cite{lent}. We find that the Berry curvature is strongly concentrated in the vicinity of two $X$ points (see Fig.6), which justifies our 
analysis of band topology based on $k \cdot p$ theory.  

\begin{figure}[tbp]
\includegraphics[width=5cm]{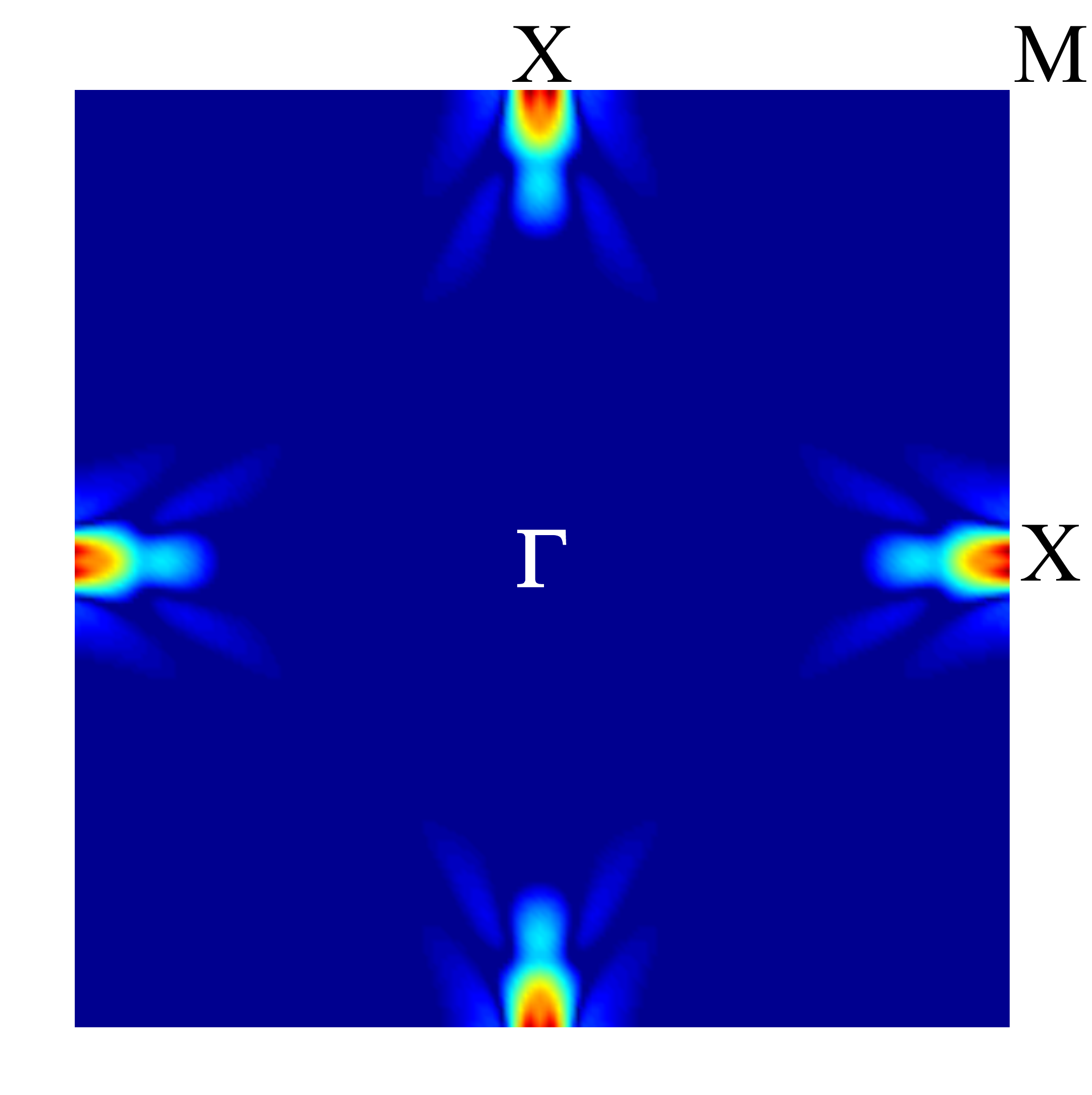}
\caption{{\bf Berry Curvature.} Plot of Berry curvature associated with $+i$ mirror eigenstates of a 5-layer SnTe film over the 2D Brillouin zone.  The $-i$ mirror eigenstates have opposite Berry curvatures. This makes the total Chern number zero, but the mirror Chern number $|N_M|=2$. }
\end{figure}

We now comment on the sign of mirror Chern number. To define the sign of $N_M$ requires first choosing an orientation for the mirror plane. This is 
because (i) mirror operation is defined as spatial inversion combined with a two-fold rotation with respect to an axis normal to the mirror plane; 
and (ii) for spin-$1/2$ particle,  the eigenvalue of two-fold rotation $\pm i$ changes sign when the axis is reversed.  Therefore the 
absolute value of mirror Chern number  {\it per se} is only meaningful after specifying the orientation of the mirror plane, and $N_M$ changes sign when 
one changes the choice of orientation.  
Nonetheless, the {\it relative} sign of mirror Chern number and electron's spin component 
along the mirror axis is an unambiguous and measurable:  it determines the directionality of spin current at the edge. 
We leave a detailed study of the sign of mirror Chern number in TCI films to a future work.      
 
 \subsection{III. $k\cdot p$ theory for edge states}
We now derive the $k\cdot p$ theory of the gapless edge states, in order to analyze how they are affected by external fields.  Note that above, the $x$ direction is actually (110) with respect to crystal axes.  Let us now choose $x$ to be (100) and analyze the gapless edge states with crossings at $\pm k_0$ along this axis.  

The two symmetries that fix a single valley are $M_z$ and $M_x \Theta$, and we choose the representation
\beq
M_z&=&-is_z \\
M_x \Theta &=& is_z K.
\eeq

The following table denotes how the electric field $E$ (in the $z$ direction) and magnetic field $(B_x, B_y, B_z)$ transforms under the two symmetries:

\begin{table}[ht]
\centering
\begin{tabular}{c | c c}
 & $M_z$ & $M_x\Theta$ \\
\hline
$E$ & $-$ & $+$\\
$B_x$ & $-$& $-$\\
$B_y$ & $-$& $+$\\
$B_z$ & $+$ & $+$\\
\hline
\end{tabular}
\end{table}

The most general $k\cdot p$ Hamiltonian compatible with the above is
\beq
H_{k_0} (k) &=& vks_z + m_{E} s_y + m_{B_{||}} s_x + m_{B_\perp} s_y \nonumber\\
&+& E_{B_z} + E'_{B_z} s_z  
\eeq
where $v$ is velocity of edge states, $m_E, m_{B_{||}}, m_{B_\perp}, E_{B_z}$ and $ E'_{B_z}$ are linearly proportional to the corresponding electric/magnetic field.  
Using time-reversal symmetry, 
we obtain the $k\cdot p$ Hamiltonian for the other valley at $- k_0$: 
\beq
H_{-k_0} (k) &=& vks_z - m_{E} s_y + m_{B_{||}} s_x + m_{B_\perp} s_y \nonumber\\
&-& E_{B_z} + E'_{B_z} s_z  
\eeq
By diagonalizing Eq.(14) and (15), we derive the field-induced gap  (4) and (5) at the two valleys.


{\it Acknowlegement}: We thank Yoichi Ando and Andrea Young for helpful comments and suggestions.  This work is supported by the U.S. Department of Energy, Office of Basic Energy Sciences, Division of Materials Sciences and Engineering under Award DE-SC0010526. T.H. acknowledges support under NSF Graduate Research Fellowship No. 0645960. J.L and W.D. acknowledge support from the Ministry of Science and Technology of China (Grant Nos.  2011CB921901 and 2011CB606405) and the National Natural Science Foundation of China (Grant No. 11074139). 
P.W. and J. S. M. would like to thank support from the MIT MRSEC through the MRSEC Program of the NSF under award number DMR-0819762, as well as NSF DMR grants 1207469 and ONR grant N00014-13-1-0301.

{\it Author Contributions}: JL performed band structure and mirror Chern number calculations. TH performed theoretical analysis with contributions from JL.   
TH and LF wrote the manuscript with contributions from all authors. 
LF conceived and supervised the project. All correspondence should be addressed to L.F.

\bibliographystyle{apsrev}

\begin{thebibliography}{10}

\bibitem{fukane}
L. Fu and C. L. Kane, Phys. Rev. B \textbf{76}, 045302 (2007).

\bibitem{teofukane}
J. Y. C. Teo, L. Fu and C. L. Kane, Phys. Rev. B {\bf 78}, 045426 (2008).


\bibitem{kanehasan}
M. Z. Hasan and C. L. Kane, Rev. Mod. Phys. {\bf 82}, 3045 (2010).

\bibitem{zhangreview}
X. L. Qi and S. C. Zhang, Rev. Mod. Phys. {\bf 83}, 1057 (2011).

\bibitem{moore}
J. E. Moore, Nature {\bf 464}, 194 (2010).


\bibitem{fu}
L. Fu, Phys. Rev. Lett. \textbf{106}, 106802 (2011).


\bibitem{hsieh}
T. H. Hsieh, H. Lin, J. Liu, W. Duan, A. Bansil and L. Fu, Nat. Commun. {\bf 3}, 982 (2012).


\bibitem{ando}
Y. Tanaka, Zhi Ren, T. Sato, K. Nakayama, S. Souma, T. Takahashi,	 K. Segawa and Y. Ando, Nat. Phys. {\bf 8}, 800 (2012).

\bibitem{poland}
P. Dziawa, B. J. Kowalski, K. Dybko, R. Buczko, A. Szczerbakow, M. Szot, E. Lusakowska, T. Balasubramanian, B. M. Wojek, M. H. Berntsen, O. Tjernberg and 
T. Story, Nat. Mat. {\bf 11}, 1023 (2012).

\bibitem{hasan}
S. Y. Xu, C. Liu, N. Alidoust, M. Neupane,	 D. Qian, I. Belopolski, J. D. Denlinger, Y. J. Wang,	 H. Lin,	 L. A. Wray,	 G. Landolt,	 B. Slomski,	 J.H. Dil, A. Marcinkova,	 E. Morosan,	 Q. Gibson,	 R. Sankar,	 F.C. Chou,	 R.J. Cava,	 A. Bansil and M.Z. Hasan,  Nat. Commun. {\bf 3}, 1192 (2012).

\bibitem{mong}
R. S. K. Mong, A. M. Essin, and J. E. Moore, Phys. Rev. B {\bf 81}, 245209 (2010).


\bibitem{vidya}
Yoshinori Okada, Maksym Serbyn,  Hsin Lin, Daniel Walkup, Wenwen Zhou, Chetan Dhital, Madhab Neupane, Suyang Xu, Yung Jui Wang, R. Sankar, Fangcheng Chou, 
Arun Bansil, M. Zahid Hasan, Stephen D. Wilson, Liang Fu, Vidya Madhavan, Science, {\bf 341}, 1496 (2013).

\bibitem{murakami}
R. Takahashi and S. Murakami, Phys. Rev. Lett. {\bf 107},166805 (2011).

\bibitem{fiete}
M. Kargarian and G. A. Fiete, Phys. Rev. Lett. {\bf 110}, 156403 (2013).

\bibitem{yao}
C. -K. Chiu, H. Yao and S. Ryu, Phys. Rev. B {\bf 88}, 075142 (2013). 

\bibitem{furusaki}
T. Morimoto and A. Furusaki, Phys. Rev. B {\bf 88}, 125129 (2013).

\bibitem{sun}
M. Ye, J. W. Allen and K. Sun, arXiv:1307.7191 

\bibitem{dai}
Hongming Weng, Jianzhou Zhao, Zhijun Wang, Zhong Fang, Xi Dai, arXiv:1308.5607

\bibitem{lee}
D. A. Abanin, P. A. Lee and L. S. Levitov,  Phys. Rev. Lett. {\bf 96}, 176803 (2006).

\bibitem{young}
A. F. Young, J. D. Sanchez-Yamagishi, B. Hunt, S. H. Choi, K. Watanabe, T. Taniguchi, R. C. Ashoori, P. Jarillo-Herrero, arXiv:1307.5104

\bibitem{kim}
P. Maher,	 C. R. Dean,	 A. F. Young,	 T. Taniguchi,	 K. Watanabe,	 K. L. Shepard,	 J. Hone and P. Kim, Nat. Phys. {\bf 9}, 154 (2013).

\bibitem{km}
C. L. Kane and E. J. Mele, Phys. Rev. Lett. \textbf{95}, 226801 (2005); {\it ibid}, \textbf{95}, 146802 (2005).


\bibitem{mbe1}
G. Bauer and G. Springholz,  Phys. Stat. Sol. (b), {\bf 244}, 2752 (2007).

\bibitem{mbe2}
E. Abramof, S. O. Ferreira, P. H. O. Rappl, H. Closs, and I. N. Bandeira, J. Appl. Phys. {\bf 82}, 2405 (1997).

\bibitem{mbe3}
A. Ishida, T. Yamada, T. Tsuchiya, Y. Inoue, S. Takaoka, and T. Kita, Appl. Phys. Lett {\bf 95}, 122106 (2009).



\bibitem{kp}
J. Liu, W. Duan and L. Fu, arXiv:1304.0430 (Phys. Rev. B in press)

\bibitem{fang}
C. Fang, M. J. Gilbert, S. -Y. Xu, B. A. Bernevig and M. Z. Hasan, Phys. Rev. B {\bf 88}, 125141 (2013).


\bibitem{hsin}
Y. J. Wang, W.-F. Tsai, H. Lin, S. -Y. Xu, M. Neupane, M. Z. Hasan, A. Bansil, Phys. Rev. B {\bf 87}, 235317 (2013).

\bibitem{kacman}
S. Safaei, P. Kacman and R. Buczko, Phys. Rev. B {\bf 88}, 045305 (2013).


\bibitem{lent}
C. S. Lent, M. A. Bowen, J. D. Dow, R. S. Allgaier, O. F. Sankey and E. S. Ho, Superlattices Microstruct. {\bf 2}, 491 (1986).



\bibitem{gf}
M P Lopez Sancho, J M Lopez Sancho and J Rubio, J. Phys. F: Met. Phys. \textbf{15} (1985).

\bibitem{du}
L. Du, I. Knez, G. Sullivan and R. Du, arXiv:1306.1925

\bibitem{onoff}
Murong Lang, Liang He, Xufeng Kou, Pramey Upadhyaya, Yabin Fan, Hao Chu , Ying Jiang , Jens H. Bardarson, Wanjun Jiang, 
Eun Sang Choi, Yong Wang , Nai-Chang Yeh , Joel Moore, and Kang L. Wang, Nano. Lett. {\bf 13}, 48 (2013).  

\bibitem{review}
D. Khokhlov, ``Lead Chalcogenides: Physics and Applications'', CRC Press (2002).

\bibitem{qw1}
A. Ishida, M. Aoki, and H. Fujiyasu, J. Appl. Phys. {\bf 58}, 1901 (1985).

\bibitem{qw2}
E. I. Rogacheva, O. N. Nashchekina, A. V. Meriuts, S. G. Lyubchenko, M. S. Dresselhaus and G. Dresselhaus,
Appl. Phys. Lett {\bf 86}, 063103 (2005).

\bibitem{qw3}
A. A. Taskin, S. Sasaki, K. Segawa, and Y. Ando, arXiv:1305.2470

\bibitem{ballistic}
G. Grabecki, J. Wrobel, T. Dietl, E. Janik, M. Aleszkiewicz, E. Papis, E. Kaminska, A. Piotrowska, G. Springholz, G. Bauer,  Physica E, {\bf 34}, 560. (2006)

\bibitem{bernevig}
C. Fang, M. J. Gilbert, and B. A. Bernevig, arXiv:1306.0888. 

\bibitem{kanerecent}
Fan Zhang, Xiao Li, Ji Feng, C. L. Kane, and E. J. Mele, arXiv:1309.7682.

\bibitem{fangz} 
Zhong Fang, Naoto Nagaosa, Kei S. Takahashi, Atsushi Asamitsu, Roland Mathieu, Takeshi Ogasawara, Hiroyuki Yamada, 
Masashi Kawasaki, Yoshinori Tokura, Kiyoyuki Terakura, Science {\bf 302}, 92 (2003).



\end{thebibliography}

\newpage

\end{document}